# A Risk Mitigation Model of Monetary Ecosystem with Stablecoins

Hongzhe Wen, Washington University in St. Louis (hongzhe.w@wustl.edu)

R.S.M. Lau, The Hong Kong University of Science and Technology (rlau@ust.hk)

**ABSTRACT**

Stablecoins have emerged as a significant component of global financial infrastructure, with aggregate market capitalization surpassing USD250 billion in 2025. Their increasing integration into payment and settlement systems has simultaneously introduced novel channels of systemic exposure, particularly liquidity risk during periods of market stress. This study develops a hybrid monetary architecture that embeds fiat-backed stablecoins within a central bank–anchored framework to structurally mitigate liquidity fragility. The proposed model combines 100% reserve backing, interoperable redemption rails, and standing liquidity facilities to guarantee instant convertibility at par. Using the 2023 SVB USDC de-peg event as a calibrated stress scenario, we demonstrate that this architecture reduces peak peg deviations, shortens stress persistence, and stabilizes redemption queues under high redemption intensity. By integrating liquidity backstops and eliminating maturity-transformation channels, the framework addresses run dynamics ex ante rather than through ad hoc intervention. These findings provide empirical and theoretical support for a hybrid stablecoin–CBDC architecture that enhances systemic resilience, preserves monetary integrity, and establishes a credible pathway for stablecoin integration into regulated financial systems.

# A Risk Mitigation Model of Monetary Ecosystem with Stablecoins

## 1 Introduction

In July 2025, the Guiding and Establishing National Innovation for US-Stablecoins (GENIUS) Act was signed into law in the United States. The Act helps create a significant regulatory framework for stablecoins, bringing greater clarity and attracting more issuers, including traditional financial institutions, to participate in the business of digital assets. Given the dominant position of U.S. dollar in the world, the Act will certainly create a tremendous impact on the existing global financial markets and monetary ecosystems.

Stablecoins enable near-instant global payments and transactions, providing a haven from volatile assets (e.g., Bitcoin), and offer access to the stability of U.S. dollar without a need of bank account and identity verification. Their rising popularity comes not from speculative trading activities but from pragmatic adoption in emerging markets, consumer spending needs, international commerce, and decentralized finance (DeFi).

The integration of stablecoins in the monetary ecosystem will certainly play an important role to modernize the existing payment systems and create a long lasting, significant impact of traditional banking functions. Such integration also places stablecoins at the center of controversies and therefore lead to the urgency of introducing a new hybrid monetary ecosystem, where public and private money interact in unprecedented ways. This paper examines the stabilities of stablecoins and introduces a hybrid monetary ecosystem in which stablecoins and fiat currencies coexist and integrate.

As financial institutions begin utilizing stablecoins for facilitating faster and programmable payments, they will be exposed to new risks beyond what have been identified in their traditional industry practices, including compliance, cybersecurity, operations, liquidity, etc. In particular, liquidity risk could be the major concern for all stablecoin stakeholders, including central banks, issuers, and users. Although stablecoins are designed to maintain a stable value by pegging their price to a reserve asset, such as fiat currency or commodities, their value could fluctuate during a financial crisis such as a bank run.

In this study, we will describe the emerging monetary ecosystem, how prominent stablecoins (e.g., USDT and USDC) will work alongside with the traditional U.S. dollar and central bank digital currencies (CBDCs). We will then propose a risk mitigation model to deal with the liquidity risk to ensure stablecoins do indeed maintain a stable value.

## 2 Related work

Since Bitcoin was proposed in 2008 as the first digital cryptocurrency, there have been widespread adoptions of blockchain-based (or distributed ledger technology, DLT-based) digital currencies issued by private companies for facilitating payment transactions without relying on any trusted third parties such as banks [1]. As more transactions are processed online, digital currencies have triggered lots of discussions about the future of our financial systems, especially in the areas of displacing existing financial intermediaries (functions traditionally performed by banks) and the needed regulatory oversights.

### 2.1 Digital currencies

Efficient and reliable monetary ecosystems are essential to facilitate all kinds of trade and everyday financial transactions. Nowadays, these ecosystems consist of a complex network of interactions among traditional financial institutions, central banks, digital currencies, and emerging technologies like blockchain and open finance platforms. Forms of money commonly used in our society may include cash, deposits, precious metals, reserve, as well as different forms of digital currencies [2].

Unlike banknotes and coins, digital currencies exist only in digital or electronic form, and include cryptocurrencies (like Bitcoin), virtual currencies (often used in online gaming environment to facilitate user transactions), CBDCs, and stablecoins. When digital currencies are issued by private companies, they are not considered a legal tender. Compared to traditional cash money, digital currency is more convenient and cost efficient for the users, especially in cross-border transactions and online payments.

Use of digital currency and online banking for financial transactions are quite different from each other. Most traditional banks nowadays offer online banking, which allows users to perform online transactions using traditional fiat currency and is subject to government regulations and banking laws. In contrast, digital currency functions independently of traditional banking systems and government restrictions. Their transactions are verified by a decentralized network, and their value is determined by market demand. While online banking is limited to traditional banking services, digital currency offers a broader range of potential applications at a much lower cost and faster time.

### 2.2 Central banks on the defense

CBDC is a digital version of a country's fiat money. As the only authorities of issuing fiat money, central banks (or destinated financial authorities) around the world have considered the issuance of their own CBDC a top priority. A 2024 study conducted by the Bank for International Settlements (BIS) shows that 94% of central banks worldwide are engaged in CBDC projects [3]. CBDCs represent an effort by governments to regulate the digital currency sector and to preserve the relevance of the traditional banking system.

As one of the pioneers of CBDC, China started research in 2014 on fiat digital currencies. Since 2020, People's Bank of China (the central bank in China) has been piloting its retail CBDC, commonly called the e-CNY, in 23 provinces and major cities using a two-layer distribution network of commercial banks and payment platforms. By mid-2024, transaction volumes of e-CNY exceeded 7 trillion CNY (about 1 trillion USD) across over 260 million wallet users, making it the largest CBDC in the world. Key features of e-CNY include offline peer-to-peer transfers, bank-backed wallet interchangeability, and programmable conditional payments, offering a real-world testbed for privacy, scalability, and cross-border use cases [4].

As the oldest central bank in the world, Bank of England also questioned what kind of money and payments would be needed to support an increasingly digital economy while observing cash as the means of payment declined from 63% in 2006 to 28% in 2018 [5]. As a result, the world's first CBDC, RSCoin was introduced in 2015 with digital ledger cryptography, making it tamper-proof and resistant to counterfeiting.

While blockchain is the technology of choice for most digital payment methods, no central bank has ever implemented a blockchain-based CBDC, citing many challenges to be solved such as performance, scalability, cross-chain interoperability and usage scenarios [6]. Another study explores how CBDCs could affect the stability of the financial system, when people shift funds from bank deposits to CBDCs or reduce their reliance on banks for money creation and credit provision as well as indirectly amplify credit-related risks during economic stress [7].

**2.3 The rise of stablecoins**

Digital currencies (other than CBDCs) generally have limited functionality as mainstream currencies, primarily due to their high volatility from price fluctuations, making them unsuitable for use as a store of value or medium of exchange. Traditional currencies are issued and backed by central banks, while most digital currencies lack such backing, impacting their stability and credibility, and hindering public willingness to accept them. Consequently, the concept of stablecoins has emerged with their value pegged to certain financial assets, most commonly a fiat currency, in particular, U.S. dollar.

Stablecoins are designed to maintain a constant value relative to certain financial assets, typically currencies, e.g., a 1:1 U.S. dollar peg. They have rapidly evolved into a "killer app" bridging the crypto ecosystem and blockchain technology with traditional financial markets and real-world applications, especially in cross-border payments and remittances. By mid-2025, the total market capitalization of stablecoins has grown to $250 billion [8], with about 99% of the stablecoins pegged to U.S. dollar [9]. The usage and impact of stablecoins are growing rapidly, with transaction volumes surpassing that of Visa and Mastercard combined in 2024 [10].

The actual adoption of stablecoins accelerated in 2017 after USDT was introduced as a U.S. dollar substitute [11]. Since then, industry practitioners and academic researchers started to discuss stability (or volatility) of different stablecoins and their impacts on the existing monetary systems. For example, while some researchers propose a framework to test the absolute and relative stability of stablecoins [12], others assess the average volatility of different stablecoin designs [13, 14]. There are also studies that compare stablecoin volatility to determine if there is any connection to Bitcoin volatility [15].

While stablecoins can enhance efficiency and open new possibilities in cross-border payments and securities markets, the Bank for International Settlements (BIS) warns that they will pose a risk to financial stability and monetary sovereignty without adequate regulation [16]. Specifically, that BIS report considers stablecoins perform poorly on singleness and elasticity, lacking the settlement function provided by the central banks, trading at varying exchange rates, and imposing a cash-in-advance constraint. There is an inherent tension between stablecoins' promise to always deliver par convertibility (i.e., be truly stable) and the need for a profitable business model that involves liquidity or credit risk.

## 2.4 Competition between CBDCs and stablecoins

The rise of cryptocurrencies and stablecoins have been seen as a threat to fiat currencies and the overall financial stability. The topic has been controversial as the impact could be dependent on a country's economic condition and other factors. Some argue there is a competition between national fiat currencies, cryptocurrencies, and CBDCs, and they find that countries with strong but non-dominant currencies are incentivized to launch CBDCs first, while the weakest economies may adopt cryptocurrencies [17].

However, many academic researchers argue that any cooperation between private companies and government authorities in digital payment does not necessarily create any competition [18]. The same paper also finds that widespread user adoption is necessary for the introduction and successful adoption of CBDCs. At the same time CBDCs must be positioned effectively alongside other digital payment options, like debit cards and stablecoins, to gain traction for widespread adoption.

An International Monetary Fund's study attempts to quantify the extent of projected adoption with a structural, choice-theoretic model incorporating banks, the central bank, the payment system, and the economic environment [19]. The study also finds that CBDC shares in total money circulation could reach 5%-25% in the United States and 1%-20% in the Euro area if the CBDC is remunerated at policy rates and perceived as "deposit-like" (with interest).

There are very few empirical studies on CBDC adoption reported to date. One exception is the study of the launch of the China's CBDC [20], which show that the adoption by individuals remain underwhelming, especially when comparing to other private e-payment methods such as Alipay and WeChat Pay, etc. The study suggests key insights that CBDC is not competitive relative existing digital payment methods and significant government resources are needed to incentivize the adoption.

If stablecoins follow the lead of other digital payment methods over CBDC, we expect to see a significant structural change of the existing financial systems based on trusted intermediaries, impacting commercial and retail banks. There are concerns that digital money such as stablecoins could substitute common bank deposits leading to financial disintermediation and instability under stress. If a country decides to adopt CBDC, industry practitioners and academic researchers generally do not recommend using a one-layer architecture, which the central bank directly manages the CBDC infrastructure. Instead, a two-layer architecture through a public-private partnership should be adopted.

**2.5 How does the GENIUS Act change the future of CBDCs?**

When the GENIUS Act was signed into law in July 2025, it marked a pivotal shift in the digital currency landscape, especially in how the United States treats stablecoins compared to CBDCs. It is clear that the United States will not issue its retail CBDCs and thus, a green light is given to the private sector to issue their own stablecoins, subject to the government oversights. While CBDCs are being sidelined in the United States, the implications to other countries are undeniable that private-sector digital currencies will drive future innovation, payments, and global competitiveness.

The Act creates a federal regulatory framework for stablecoin issuers to follow. For example, issuers must fully back each stablecoin with a 1:1 ratio of security assets, and reserve assets must be limited to highly liquid investments such as cash and Treasury bills. Issuers must disclose their reserve holdings monthly and comply with anti-money laundering regulations. If the market capitalization of a stablecoin exceeds $50 billion, the issuer must submit annual audited financial statements. The Act also protects consumer rights, granting holders priority access to reserve assets in the event of the bankruptcy of a stablecoin issuer.

As almost all of the stablecoins are pegged to U.S. dollar in 2025, it is considered a major milestone for stablecoins when the United States finally establishes a federal regulatory framework for stablecoins through the GENIUS Act. Given the influence of U.S. dollar and the size of its financial markets, the Act will no doubt serve as a reference for other countries to adopt similar measures in their future stablecoin-related legislation.

This mechanism effectively links the issuance of stablecoins to demand for U.S. dollar and Treasury bills, making global stablecoin users indirectly "retail buyers" of U.S. Treasury bills. If stablecoin issuance expands from the current $200 billion to the projected

$2 trillion, global funds will provide the United States with a large and continuous source of funding, which explains the active push by the United States for stablecoin legislation.

Meanwhile, facing more challenges from stablecoins in cross-border transactions and settlements, SWIFT, which functions worldwide as a secure messaging service connecting 11,000 financial institutions and processing billons of messages annually for interbank settlements, has begun testing on-chain payments and messaging using the blockchain technology [21]. This represents a significant transformation for international payments, which largely remain dependent on intermediaries and legacy infrastructure.

## 3 Development milestones of stablecoins

Since stablecoins are backed by other financial assets. As expected, these types of stablecoin are subject to frequent price fluctuation and render them less common in use. On the other hand, algorithmic stablecoin (e.g., TerraUSD or UST) is fundamentally different because it is not backed by any financial assets and relies entirely on algorithms to adjust its issuance value by balancing the supply and demand. Any imbalance of supply and demand can fluctuate its value.

For example, the value of UST collapsed in May 2022, likely triggered by an attack on its liquidity pool and a flaw in its algorithmic peg mechanism, and caused billions of dollars in loss to investors. It is now just a legacy token of the failed ecosystem, with a value of only $0.014 in August 2025.

Stablecoins are primarily used for trading crypto assets, settling payments in goods and services, insulating against local currency instability, and sending payments across borders. BitUSD is commonly known as the first stablecoin, released in 2014 as a crypto-backed stablecoin. It is collateralized with a limited amount of reserve from BitShares, making its price volatile especially when there is an imbalance between supply and demand of BitShares. Eventually, BitUSD lost its 1 to 1 parity with U.S. dollar in 2018 and could never recover since then.

Other early adopters of the stablecoin concept included NuBits and UST but they also failed for a lack of truly stable reserve assets backing [22]. Despite some early failures and setbacks, many global financial service providers have come up with a better model or mechanism to stabilize t heir stablecoin values, resulting in much wider acceptance.

### 3.1 USDT/USDC – Asset/Fiat collateralized stablecoin model

In February 2025, the market capitalization of USDT and USDC accounted for 64% and 25% of the stablecoin market share, respectively [23]. Tether (USDT), launched in 2014 by Tether Limited, still stands as the most widely adopted stablecoin, maintaining quite steadily a 1:1 peg with U.S. dollar. USDT's prominent place in the cryptocurrency market is recognized by its substantial market capitalization and trading volume.

USD Coin (USDC) shows another leading example of a fiat-collateralized stablecoin, launched in 2018 by the Centre Consortium (led by Circle and Coinbase). USDC is designed to maintain a 1:1 peg to U.S. dollar by holding dollar-denominated reserves backing every coin in circulation [24]. For each USDC issued, there is supposed to be one U.S. dollar (or equivalent safe asset like a Treasury bill) held in custody [25].

The stability of USDC's peg comes from arbitrage and trust: If USDC's market price ever falls below $1, traders can buy it cheap and redeem for $1 of fiat money, pushing the price back up; if it rises above $1, they can issue or release reserves to bring it down [26]. If, under extreme conditions, sustained downward pressure on the stablecoin happens, USDC's issuer would liquidate its reserves to meet redemptions.

## 3.2 DAI – Algorithmic stablecoin model

Other than USDT as mentioned earlier, DAI represents another example of decentralized, crypto-collateralized stablecoin model. DAI is issued by the MakerDAO protocol and is soft pegged to $1 through an over-collateralization and feedback mechanism rather than fiat reserves [27]. Users generate DAI by locking up volatile crypto assets (such as Ethereum, and now various others, including USDC itself) in smart contracts called Maker vaults [28]. The system is designed to ensure the collateral value exceeds the amount of token that DAI issued (typically a minimum of 150% of DAI's value in collateral). If DAI trades below $1, the protocol can raise stability fees or encourage collateral auctions to reduce DAI supply. If above $1, it can lower fees or allow more DAI to be minted, aiming to equilibrate supply and demand [29].

## 3.3 Regulations on stablecoins

Stablecoins are cryptocurrencies with (supposedly) a stable value that are pegged to fiat currencies such as U.S. dollar, euro, and yen, or to specific assets such as gold. The core design of stablecoins is to reduce price volatility and provide a stable store of value and medium of exchange. From the perspective of actual users and the general public (potential users), there are many risks associated with stablecoins, including:
- Insufficient collateral: The reserves supporting fiat currency stablecoins may not be enough to meet the demand when more users redeem at the same time, similar to a bank run.
- Intermediaries may not be trustworthy: The essence of stablecoins is primarily determined by the issuer (institution) that controls the money and conducts/blocks the transactions.
- Systemic risk: If a single stablecoin is widely used on payment platforms, it will cause a systemic risk (or disruption of economic activities) when too much power is concentrated in that issuer.
- Illegal activities: Stablecoins can be misused and facilitate illegal activities, such as money laundering and terrorist funding.

In response to a call for legitimizing the stablecoins while providing more consumer protection, strengthening financial stability, and curbing illegal activities, all the major financial hubs have moved from "wait-and-see" to full-fledged regulatory regimes for fiat-backed stablecoins. From 2023 to 2025, new legislations in Singapore, European Union, Switzerland, United Aarab Emirates, Hong Kong, and the United States have helped legitimize stablecoins and provided a framework for monitoring their operations.

Collectively, these frameworks mark a decisive move toward integrating fiat-backed stablecoins into mainstream prudential supervision. By contextualizing crypto as a truly financial innovation, monetary authorities around the world have helped remove many uncertainties and obstacles for the public to accept stablecoins. Yet, due to the many divergences in different legislations, some practical frictions remain, such as passport constraints, collateral-location requirements, and multiple examination schedules, that any cross-border hybrid monetary architecture must address through interoperability standards or multi-licensing strategies.

## 4 Emerging risks of stablecoins

As financial institutions begin using stablecoins to facilitate faster and programmable payments, they will need to confront a range of emerging risks in areas of compliance, cybersecurity, operations, and liquidity [30].

### 4.1 Compliance risk

Transactions in stablecoins raise considerable, complex compliance challenges to financial institutions, especially when it comes to anti-money laundering (AML) risks and to satisfy the know your customer (KYC) requirements, because stablecoins can move easily outside traditional banking networks and potentially be exploited by illicit actors.

To counter compliance risks associated with stablecoins in an effective way, financial institutions should focus on (re)designing a process to ensure customer due diligence for all stablecoin users (e.g., verifying their identities and wallet ownership) by monitoring and screening all transactions for suspicious activity. Regular compliance audits should be conducted to make sure all necessary steps are taken to adhere to the policies and industry best practices. To monitor such a large number of transactions on a continuous basis, advanced technology, such as an automated screening system, must be deployed to support the people and enable the process to achieve the compliance objectives.

### 4.2 Cybersecurity risk

Concerns of cybersecurity and blockchain infrastructure will begin to rise as financial institutions need to deal with digital wallets, private keys, and APIs that connect to blockchain networks. Proven IT protocols and blockchain security measures are necessary to protect financial institutions against threats like phishing, unauthorized transactions, and vulnerabilities in smart contracts etc. Investments in technology, controls, and talent are essential to build a secure, trusted digital asset ecosystem.

An industry best practice is to establish rigorous security protocols for all stablecoin-related operations. This includes multi-factor authentication for system access and transaction initiation and multi-person approval workflows for transfers above set thresholds to prevent a single point of failure. Another best practice is to enforce the principle of least privilege and segregation of duties, to ensure no individual can unilaterally control funds. Backup and key-recovery procedures must also be in place to prevent loss of access. In addition, transaction risk scoring tools can be deployed to identify suspicious patterns and halt them for review.

### 4.3 Operational risk

Operational risk arises from the internal processes, systems, and third parties on which stablecoin transactions rely. Unlike traditional payment systems, stablecoin operations depend on blockchain networks and external issuers, which can easily introduce new points of failure and complexity.

To ensure stablecoins are handled properly at all times, some existing processes should be redesigned to handle stablecoin payment activities to include extra steps like dual control and verification for critical tasks, end-to-end reconciliation of stablecoin transactions with internal ledgers to make sure that the amount of stablecoins held in custody matches customer balances and fiat reserve movements at all times. The redesigned process should also appropriate due diligence and oversight of any third-party service providers, ensuring they have strong uptime records and incident response commitments. All of these new measures may require additional technology investment for more reliable infrastructure and platforms to support stablecoin operations.

### 4.4 Liquidity risk

Liquidity risk refers to the ability to readily convert stablecoins to cash (or vice versa) without delay or loss of value. In normal conditions, a stablecoin may function as a cash equivalent. When the financial system is under stress or going through a crisis (e.g., a run on the stablecoin or a market panic), it is uncertain whether the token can be promptly redeemed for cash at par. This uncertainty was highlighted by the temporary de-pegging of major stablecoins during recent market turmoil.

As evidenced by the Silicon Valley Bank (SVB) incident, when a large amount of a certain stablecoins was deposited there, a bank run could effectively impact the redemption of the stablecoin and cause a de-pegging. Banks involved in stablecoin payments also face counterparty and credit risk as they rely on the stablecoin issuer to maintain adequate reserves and honor redemptions. If an issuer's reserves are incomplete, illiquid, or held with risky custodians, the stablecoin's value could falter, impacting the bank and its customers. In effect, users and banks must trust the issuer's integrity and stability.

To counteract the liquidity risk, banks will need to monitor its stablecoin exposures and ensure convertibility. Dedicated risk managers should be appointed to evaluate and monitor regularly the financial health of stablecoin issuers, tracking the inflows/outflows of stablecoins to identify potential liquidity gaps.

When banks act as the custodian for a stablecoin issuer, they should consider a liquidity stress test to find out if they could still meet the outflow demand in case that issuer temporarily suspends redemptions. They should develop a contingency plan and be ready to communicate with clients to prevent panic when a stablecoin loses its peg. Banks could consider using real-time treasury monitoring systems or integrating with the stablecoin issuer's platform to track stablecoin positions and movements. Some advanced analytics software can also be deployed to watch market indicators to suggest a stablecoin is straying from its peg, or early warning signals of stress when large volumes are exiting.

## 5 A proposed architecture of hybrid monetary ecosystem

There is a growing support for CBDCs and stablecoins (or other digital payment methods) to coexist by creating a hybrid monetary system with a two-tier structure: wholesale and retail layers. For example, digital euro and e-CNY have adopted a two-tier or intermediated architecture to prevent the disintermediation of the commercial banking sector [31]. CBDCs are issued and regulated directly by central banks, offering a stable, government-backed digital currency, and are intended for regulated payment systems, bolstering monetary stability, and potentially improving financial inclusion and more oversight. Concerns exist regarding privacy and potential government surveillance.

Meanwhile, stablecoins have already found a strong use case within decentralized finance (DeFi) by offering high interoperability and integrating quickly into private sector platforms. They face increasing regulatory scrutiny to ensure transparency and stability after high-profile collapses, such as TerraUSD in 2022 when it failed to maintain its $1 price target and triggered widespread sell-offs across the crypto space. This study will address the liquidity risk issue with a proposed hybrid monetary model.

While financial institutions can implement multiple measures to contain the liquidity risk, the most effective way is to design a system architecture that can prevent the liquidity risk rather than respond to it. To further enhance the stability of stablecoins to prevent the liquidity risk, the proposed hybrid monetary system is a two-layer architecture combining private-sector innovation at the user interface and central bank support at the core:

- Layer 1 – Core CBDC infrastructure (public): The Federal Reserve operates the wholesale/retail CBDC ledger and grants interfaces to regulated intermediaries. It supplies the default-free settlement asset while leaving retail customer relationships to the private sector.
- Layer 2 – Private tokens and deposits (private): Banks or fintech issuers tokenize deposits and circulate stablecoins that are redeemable 1:1 for CBDC. Reserve assets are held as CBDC balances or Fed-custodied Treasuries, making issuers functionally narrow banks. Interoperability is achieved through shared messaging and redemption protocols.

At a high level, regulated private entities (stablecoin issuers and banks) would issue digital USD tokens and provide payment services, while the Federal Reserve and government establish or adopt the common foundation (rules, settlement infrastructure, and backstop). As shown in Fig. 1, this structure mirrors an "intermediated" model of CBDC but extends it by treating existing fiat-backed stablecoins as a feature rather than a competitor. The design rests on several pillars as explained later.

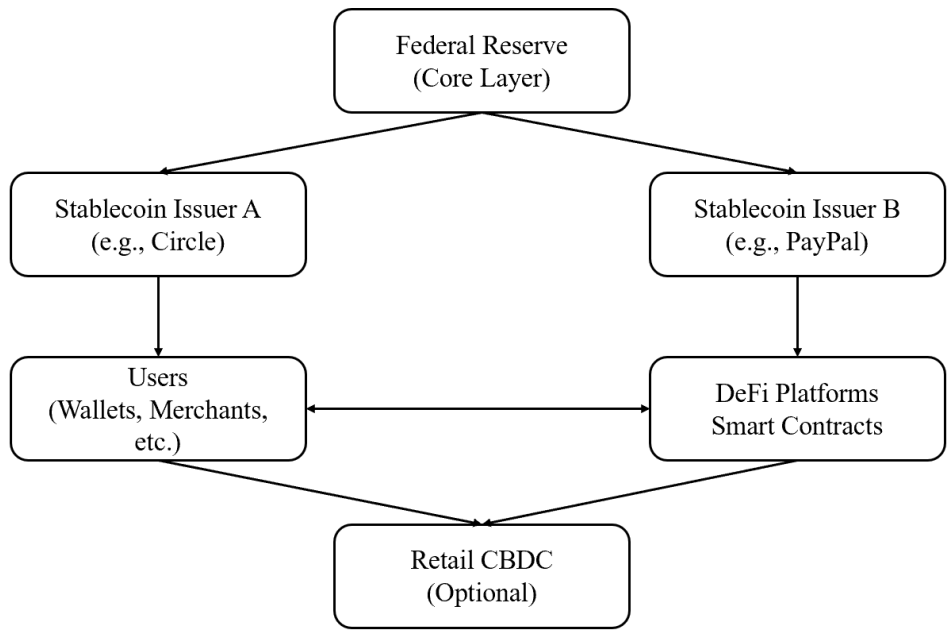

**Fig. 1** Hybrid monetary system design diagram.

## 5.1 Private digital dollar issuers

In this proposed model, fintech companies or banks will issue tokenized USD liabilities (stablecoins) that are fully backed by high-quality, liquid assets or actual fiats. Each stablecoin token represents a claim to $1 in reserves. Users can hold and transfer these tokens on various blockchain networks, or called public chains (Ethereum, Solana, etc.), enabling instant P2P payments, smart contract integration in DeFi, and more. These issuers will operate under charters/licenses that meet strict compliance requirements.

In practice, this means firms like Circle (issuer of USDC), Tether (issuer of USDT), or other non-bank payment firms (e.g., PayPal with PYUSD) could become "Permitted Payment Stablecoin Issuers" as defined in new legislation. Traditional banks could also issue their own tokenized deposits, but the model focuses on interoperable stablecoins that any user can accept, rather than each bank creating a silo stablecoin and competing with each other.

To ensure zero-delay USD conversion, issuers will connect directly into FedNow rails: any on-chain redemption request triggers a same-second FedNow credit to the user's bank account. This binding commitment eliminates common "cash-out" issues in private stablecoins today.

## 5.2 100% reserve backing (synthetic CBDC model)

To ensure uniform value and eliminate liquidity (similar to a bank run) risk, each stablecoin issuer in the system needs to hold reserves equal to 100% of their outstanding tokens in safe assets (USD cash or equivalents). The optimal arrangement, often called a "synthetic" CBDC, is for issuers to deposit these reserves directly in the Fed account.

The stablecoin effectively becomes a narrow-bank liability fully backed by central bank money. This gives the tokens or stablecoins the same creditworthiness as CBDC (since the backing funds are a claim on the Federal Reserve) while still allowing private issuance. If direct Fed access for non-banks is initially unattainable, a similar effect can be achieved by holding reserves in short-term Treasury bills or insured bank deposits under trust. In the end, the goal is to minimize liquidity risk in reserves for stablecoins.

## 5.3 Federated networks and interoperability

To prevent fragmentation and competition, the system requires interoperability among all official stablecoin tokens and with other forms of USD. A stablecoin issued by one licensed entity is fungible with another's and with traditional dollars. Achieving this requires common technical and legal standards; for example, standardized messaging formats and APIs to form bridges between different platforms, which enables transferring and swapping of tokens easily.

An interoperable design could allow a user with USDC tokens to easily send funds to someone who uses, for example, a PayPal PYUSD wallet, or pay directly into a merchant's bank account, which the underlying infrastructure would convert and route the payment, much like how different banks clear checks at par nowadays.

The Federal Reserve could facilitate this by extending its payment systems (FedNow or FedWire) to support stablecoin settlement, e.g., allowing banks to settle stablecoin redemption flows in real time. Industry consortia might develop common protocols or bridges to connect multiple blockchains, so stablecoin liquidity is not isolated on one chain.

**5.4 Role of a retail CBDC**

The model does not necessarily require the Federal Reserve to issue a retail CBDC immediately but leaves space for one as a complementary option. The Federal Reserve could proceed with a limited retail CBDC aimed at specific use cases (for example, as a public option for those who prefer holding a direct claim on the central bank) or only available when off peg happens to stablecoins. This CBDC would function alongside stablecoins: Think of it as the digital equivalent of physical cash, whereas stablecoins are more like digital demand deposits issued by private institutions.

In our hybrid design, the CBDC and stablecoins coexist and are interchangeable. A user might hold some balance in CBDC for its sovereign guarantee, and some in a private stablecoin wallet for specialized services, such as DeFi lending/borrowing or international commerce purposes. These wholesale CBDC tokens would settle instantaneously in seconds and 24/7 with almost no marginal cost, mirroring the sub-cent fee structure observed in retail CBDC pilots such as the digital euro (€0 per user transaction) and Nigeria's eNaira (0 NGN per P2P transfer).

While a retail CBDC offers consumers the same instant and near-zero-fee experience, it also raises privacy, disintermediation, and cybersecurity concerns that are less pronounced in the wholesale context. By confining CBDC tokens to the wholesale layer, banks and issuers can convert stablecoin liabilities into CBDC units on-chain without exposing end-users to those trade-offs: Preserving commercial bank funding while ensuring finality and minimal settlement risk. Ultimately, every tokenized dollar, whether retail stablecoin or wholesale CBDC, would need to clear back to a Fed liability, maintaining a single, centrally controlled monetary base that supports both innovation and financial stability.

**5.5 Programmability and DeFi integration**

A key advantage of including private stablecoins is its rich ecosystem of decentralized finance (DeFi) and programmable money. The hybrid system would support smart contract functionality for the dollar. This means stablecoins can be used in automated agreements through smart contracts, e.g., release funds on delivery of goods, yield-generating decentralized lending, or micropayment streams by the second.

By design, the stablecoin tokens would be composable with existing DeFi protocols. In fact, USDC and DAI are already deeply embedded in lending platforms, decentralized exchanges, and so forth. The Federal Reserve or any regulators need not build this functionality themselves. Instead, they supervise the private service providers who innovate in it. The result is a programmable money layer on top of the dollar, integrating fintech innovation further into the system.

For businesses, this enables cost-effective payment logic and internal circulation (like supply chain payments that execute automatically upon IoT sensor triggers, etc.), and for consumers, it means access to a global financial marketplace using dollars in new ways (borrowing/lending, trading, automated investments) without currency exchange friction.

In the United States, this hybrid system helps with maintaining USD primacy to a further step by ensuring the dollar is the default currency for value in the future digital economies. That is, by providing official support for USD stablecoins, the United States can crowd out potential competitors, and such benefits are maintained the same for all countries who adopt the system. Conversely, if no USD token were readily available, a crypto market might adopt a foreign stablecoin or even a volatile crypto as a unit of account, which ensures the stability of the hybrid system.

**5.6 Monetary integrity and parity**

Crucially, this architecture preserves monetary integrity – all dollars, whether in one's bank account, a wallet, or a smart contract, are equivalent and fully fungible. There is one monetary base (the Fed's liabilities) supporting multiple manifestations. Unlike today's situation where a stablecoin could default or de-peg independently, any loss of peg is virtually impossible in the system because of the full-reserve rule and central bank custody.

The Federal Reserve and regulators ensure that no participating stablecoin can be issued beyond its reserves or engage in risky reserve investments. If a stablecoin issuer violates rules, immediate redemption by arbitrageurs will force it back into compliance or out of the system. In effect, the stablecoins become an extension of the monetary base, not a parallel currency. This prevents a scenario where "two dollars" (say, a Federal Reserve issued CBDC dollar and a private stablecoin dollar) diverge in value or credibility, which is a risk that would undermine the dollar's singular role.

Because the Federal Reserve oversees the reserve backing, it can manage the overall money supply implications. If demand for digital dollars rises (when people convert bank deposits or cash into stablecoins), those funds end up as reserves at the Federal Reserve, which is part of M0. The Federal Reserve can offset by open market operations if needed to maintain its policy stance. Interest in reserve balances could become a tool to influence stablecoin usage (e.g., if the Federal Reserve pays interest on reserves, perhaps stablecoin issuers could pass some of it to token holders, aligning with policy rates). All of this ensures that even as technology evolves the form of money, the Federal Reserve retains ultimate control over monetary stability and liquidity.

## 6 A Use Case Example

To illustrate the architecture as mentioned earlier, here we consider a user scenario in the following hybrid system. Alice, an individual, has $500 in a bank account but wants to participate in a DeFi investment. She uses a fintech app to convert $100 into a regulated stablecoin token. Behind the scenes, the issuer takes $100 from Alice via Automated Clearing House (ACH) system from her bank and places $100 into its Fed's reserve account, minting 100 digital tokens in return, which it delivers to Alice's crypto wallet. Alice can then send some tokens to a friend or merchant with ease via a mobile blockchain wallet in seconds (domestic peer-to-peer payment) and deposits the rest into a DeFi lending platform where it earns risk-free interest programmatically. The tokenized money flow is depicted in Fig. 2.

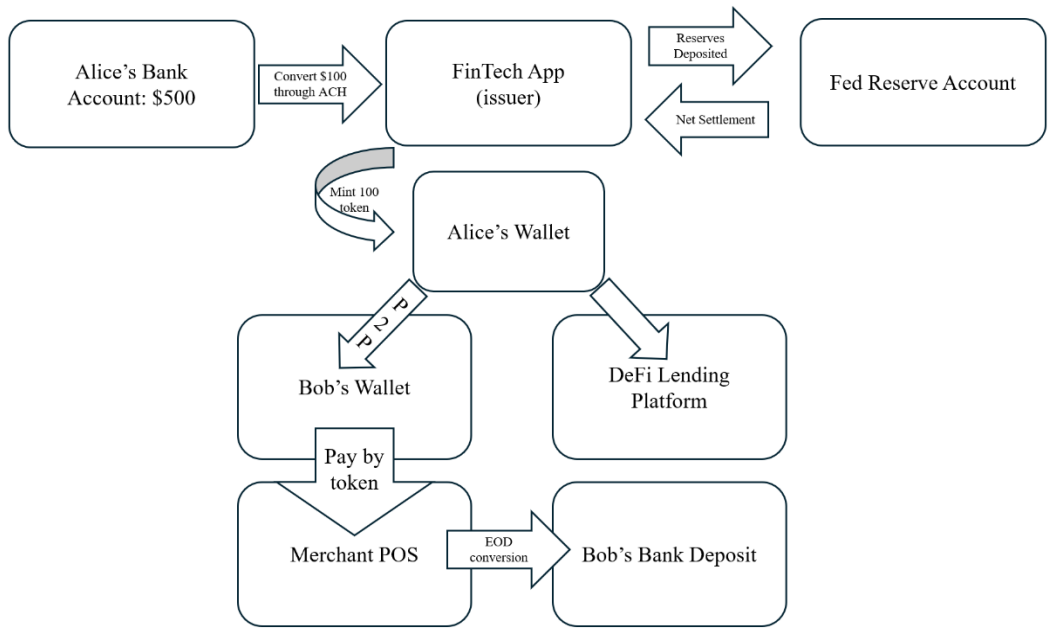

*Alice converts $100 via ACH from her bank into 100 stablecoin tokens, which are fully backed by reserves held at the Federal Reserve. She then uses her crypto wallet for peer-to-peer payments and DeFi investments, while merchants accept the tokens at POS (point-of-sale) and convert them back to USD via an EOD (end-of-day) settlement, which all reconciled with the Federal Reserve in the background.*

**Fig. 2** Tokenized money flow in the hybrid monetary system.

On the other hand, Bob, a merchant, accepts stablecoin payments from customers (Alice). His point-of-sale (POS) system can convert those tokens to his preferred stablecoin or directly to a bank deposit at the end of day through an integrated service, so there is no price risk in accepting one versus another. If Bob's business would rather keep some revenue in stablecoins, it could invest in a DeFi yield fund first, then redeem to US dollar when needed for supplier payments.

Meanwhile, in the background, the stablecoin issuers and banks cooperate to reconcile net flows through the Federal Reserve's system. If issuer A has net redemptions, it reduces its Federal Reserve's reserves accordingly, and if another issuer B sees net inflows, it then increases B's reserves to balance. The Federal Reserve's balance sheet expands or contracts in tandem, but all dollars remain within the regulated banking perimeter (either as bank deposits or reserve deposits backing tokens). In this way, everyday economic activity can use tokenized money without leaving the safety net of the central bank.

## 6.1 Comparison to the existing system

This hybrid model absorbs the pros and leaves the cons of the two extremes to create a purely public solution (the Federal Reserve issues retail CBDC to everyone and bans private stablecoins) and a laissez-faire approach (stablecoins can circulate without any special regulation or Federal Reserve's involvement). This model, while guaranteeing uniformity, would suppress private innovation and raise concerns about the government operating retail accounts. The laissez-faire approach could also lead to destabilizing runs or a "wild" situation of unregulated private monies being used under improper situations.

Note that China's e-CNY pilot program points out both the advantages and downsides of a full-retail CBDC. Its rapid user adoption demonstrates strong demand for a sovereign digital currency, yet early trials highlight trade-offs around privacy (geolocation control, etc.), interoperability with traditional bank accounts, and the high demand for powerful cyber-resilience infrastructures. These practical launch insights from e-CNY suggest that our hybrid design by emphasizing how wholesale-only and optional retail layers can be balanced to mitigate privacy and over-centralization concerns.

Our hybrid approach seeks a balance between keeping what works in the market (speed, innovation, user-driven adoption of stablecoins) and adding the Federal Reserve's imprimatur and oversight to ensure safety and unity. By formalizing stablecoins as part of the system, we also ensure inclusivity. That is, the digital dollar revolution is not limited to crypto enthusiasts but can be accessed via familiar interfaces (your bank or payment app might offer stablecoin services as well as the wallets).

The design is also extensible internationally. If other countries adopt similar frameworks (or if the United States coordinates via standard-setting bodies), cross-border payments could be simplified by bridging CBDCs and stablecoins from different jurisdictions, and a US stablecoin token could be swapped for a digital euro token through automated forex smart contracts, enabling remittances in seconds at an extremely low cost.

## 7 Could SVB bank run and USDC de-peg be prevented?

In March 2023, the crypto market was shaken when USDC broke its $1 peg due to a shock spreading from the traditional banking system. We examine what happened and how the hybrid framework could have improved performance and stability during this crisis.

### 7.1 Event background

On March 10, 2023, Silicon Valley Bank (SVB), a large US bank, collapsed and was taken over by regulators. By the end of 2022, only 6% of the deposit in SVB was insured, and the rest remained uninsured. Circle, the issuer of USDC, disclosed that $3.3 billion of USDC's reserve funds were deposited at SVB. This represented about 8% of USDC's total reserve that was potentially frozen or lost. Panicked USDC holders rushed to sell or redeem USDC. As a result, USDC's market price plunged to $0.87 on March 11. This de-peg was unprecedented for USDC, which historically fluctuated by only pennies.

Other stablecoins also reacted: DAI, with over 50% of its reserves being collateralized by USDC at the time, also dropped to around $0.90. Tether (USDT), which had no exposure to SVB, saw increased demand and traded slightly above $1 (around $1.01 to $1.02). Some trading and DeFi protocols faced chaos due to the unexpected price discrepancies of "supposed-to-be $1" tokens.

### 7.2 Resolution

The crisis was resolved on March 12 when US authorities announced that all SVB depositors would be made whole on Monday, eliminating the risk that Circle's reserves would suffer a loss from the crisis. Circle also publicly assured that USDC remained redeemable 1:1 and that any shortfall would be covered by other resources if needed. After the news, confidence restored, and USDC's price rapidly climbed back towards peg. By March 13, the price of USDC was around $0.998, and by March 15, it was $1.00 again. In total, USDC spent roughly 3 days under $0.99.

Notably, crypto exchanges and DeFi platforms had to navigate this interim period by ceasing some USDC transactions or updating their codes. Ultimately, the situation was resolved without any losses to USDC holders, but it exposed a flaw: Stablecoins are only as stable as their backing and the confidence in it. In this case, the backing was entangled with the banking system and reliant on an ad hoc government intervention to avoid a disastrous outcome.

## 7.3 The Diamond-Dybvig model and run equilibrium

Diamond-Dybvig's framework shows how banks that issue liquid deposits and invest in illiquid assets are subject to self-fulfilling runs arising from the existence of multiple equilibria. In normal times (the no-run equilibrium), depositors believe the bank is solvent and will honor withdrawals, so only those with genuine needs withdraw and the bank can meet demands. But if depositors expect others to run, it creates a panic: Everyone tries to withdraw immediately out of fear of being last in line and losing their money.

In SVB's case, we can map the events to a shift from a no-run to a run equilibrium triggered by a shock to expectations:

- Initial state (No-run equilibrium): Before March 2023, SVB depositors generally believed the bank was sound. Even though SVB had large unrealized losses, depositors did not exhibit withdrawal pressure because they expected SVB could meet withdrawals. This aligns with the Diamond–Dybvig no-run scenario: if each depositor expects others not to withdraw en masse, it is rational for them to also leave their funds deposited and avoid causing a run.
- Expectation shock: SVB's March 8 disclosure was the shock that reversed depositor expectations. The announcement of a loss and an emergency capital raise sent a signal that the bank's solvency might be impaired. This quickly became common knowledge among depositors. Large depositors and market commentators expressed concerns about SVB's solvency. In Diamond–Dybvig terms, depositors' belief shifted to "others might run" and therefore accelerated their own withdrawals." Once depositors doubted that SVB had enough liquid assets for everyone, the run equilibrium became the rational expectation. Each depositor had a strong incentive to withdraw immediately because if they waited and others ran, SVB would fail and their uninsured funds could be lost or locked up.
- Run equilibrium dynamics: As the model predicts, once a critical mass began withdrawing, it was optimal for all uninsured depositors to join the run. The fact that about 90% of SVB's deposits were uninsured greatly exacerbated this dynamic. In the Diamond–Dybvig framework, deposit insurance can eliminate the run equilibrium by assuring depositors they will get their money even if the bank fails. But at SVB, most depositors had no such guarantee – their money was at risk if SVB went under. Thus, the rational choice was "withdraw now to mitigate potential loss." Indeed, on March 9 we saw a classic coordination failure: thousands of depositors tried to withdraw approximately $42 billion in one day, far exceeding SVB's available liquidity. Banks operate on a sequential service basis, so those who ran first would withdraw at par value, while late-comers risk being stuck with whatever assets remained. This first-mover advantage is at the heart of the Diamond–Dybvig run equilibrium.

To illustrate with a simplified numerical example: Suppose a bank's long-term assets will be worth $1.00 per $1 of deposits if held to maturity (assuming rational behavior), but only $0.70 per $1 if liquidated early in a fire sale during a run. In a no-run scenario, each depositor can ultimately get $1 back (with interest) by waiting. But in a run scenario, if you wait, you might only get $0.70 (or nothing if others drain the liquidity of the bank first). Facing that prospect, even a rational depositor with no immediate need prefers to withdraw now and obtain the full nominal amount. That logic is self-fulfilling: if everyone follows the same logic, the run occurs and the bank indeed fails, confirming the expectation.

In the SVB incident, once the equilibrium shifted, existing market mechanisms were insufficient to prevent the run. Only an external intervention (the government's deposit guarantee on March 12) restored confidence by effectively removing the risk of loss for depositors. This echoes Diamond–Dybvig's prescription that deposit insurance or a lender-of-last-resort can prevent runs by guaranteeing liquidity and breaking the vicious cycle. Notably, after the authorities guaranteed all SVB deposits, withdrawal pressure subsided once loss-given-default was eliminated, since their expectations shifted back to safety.

The SVB's incident can be seen as a real-world instantiation of the Diamond–Dybvig theory of bank runs. The bank's heavy reliance on uninsured deposits meant the no-run equilibrium was fragile. This underscores that in modern banking, psychology and coordination problems can be just as lethal as bad assets. Even a fundamentally solvent bank can be brought down by a self-fulfilling run if confidence evaporates.

### 7.4 How could the hybrid system help?

The hybrid architecture does not promise that stress events will never occur, but it does change both the probability and the propagation of a crisis. To illustrate, consider how the SVB–USDC episode in 2023 would have unfolded had the hybrid system already been in force:

*Prevention of reserve risk:* Under the model, every privately issued dollar token must be collateralized with central-bank liabilities or assets that can be converted to those liabilities at face value within one business day. Had Circle's $3.3 billion reserve been placed at, or swept daily into, an omnibus account at the Federal Reserve, SVB's failure would still have occurred, but the credit exposure of USDC holders to SVB would have been zero. In other words, the framework eliminates the specific maturity-transformation channel that triggered the de-peg. It does not abolish all sources of volatility (e.g., market sentiment), but it removes the direct link between a commercial-bank insolvency and a stablecoin reserve shortfall.

*Lender of last resort and liquidity:* Full-reserve custody does not preclude temporary liquidity mismatches. For example, if a portion of collateral is held in T-bills that must be settled on a T+1 base, the model can extend standing collateralized credit lines that are parallel to the discount window to licensed issuers. If redemptions surge faster than securities can be settled, the issuer can pledge those same securities to the Federal Reserve and receive immediate settlement balances. This converts what could have become a redemption queue into an operational-timing issue resolved intraday.

*Transparency and trust:* Another improvement is that under a regulated hybrid system, real-time transparency of reserves might be available. If users and traders could see (via publicly verifiable means) that USDC had full backing in safe assets, they might not have panicked as much. In 2023, there was uncertainty and news-driven fear since most USDC holders were unclear about the details of the reserve. The hybrid model could employ blockchain proofs or Fed-published data to show reserve status continuously to avoid any panic caused by uncertainty.

*Federal Reserve/Treasury as the stabilizer:* Under a closer public-private partnership, authorities might directly stabilize a critical stablecoin if needed. Consider if the SVB news had not been resolved so quickly, perhaps regulators might have let uninsured depositors take a haircut, USDC could have remained de-pegged. In a hybrid system scenario, because of the potential systemic importance, the Federal Reserve and Treasury in coordination with the issuer could act. For example, the Federal Reserve could temporarily swap any reserves that SVB held with its own funds to make the stablecoin safe (effectively a mini rescue for the stablecoin holders, using the issuer's other assets as collateral).

*Interoperability and redundancy:* In our model, hypothetically, if USDC had a problem, people could convert 1:1 into an alternate token through the network, rather than sell at a discount. This mutual fungibility acts as a pressure valve and no single failure would strand users without parity conversion options. It is similar to deposit insurance in banking, in which, even if a bank fails, depositors know their money is safe and effectively interchangeable with money at any other bank up to insurance limits. The hybrid currency network would extend that assurance across digital dollars.

### 7.5 Cross-border AML/KYC

The hybrid network spans jurisdictions that apply the Financial Action Task Force (FATF) Standards at different levels of maturity. The updated FATF Recommendation in June 2025 now requires originator and beneficiary data (name, address, date of birth) to accompany all peer-to-peer cross-border transfers above 1,000 EUR (or USD), whether processed by banks, virtual asset service providers (VASPs), or token-transfer protocols [32].

Parallel FATF guidance urges supervisors to accelerate licensing of offshore VASPs and to close the stablecoin "gaps", highlighted in the April 2025 Virtual-Assets Targeted Update. All licensed issuers and access gateways participate in an InterVASP-compliant secure channel. For on-chain transfers between self-hosted wallets, gateways attach a hashed Travel-Rule envelope that can be decrypted on regulator demand, meeting the full-information chain test without revealing personal data on-chain.

End-users bind a World Wide Web Consortium (W3C) verifiable credential, issued by a FATF-regulated entity, to their wallet address. When they switch service providers cross-border, the credential is re-used, lowering onboarding friction while preserving the one-time know your customers (KYC) audit trail.

The European Union's General Data Protection Regulation (GDPR), Singapore's Personal Data Protection Act (PDPA), and Switzerland's Federal Act on Data Protection (FADP) all restrict export of personally identifiable information. By encrypting, or zero-knowledge-proofing, the minimum Travel-Rule fields, the network can route compliance packets through regional "compliance enclaves" that keep raw identity data inside the originating jurisdiction. Only salted hash travels with the payment instruction, satisfying both FATF transparency and local privacy statutes.

FATF's June 2025 report notes that stablecoins now dominate on-chain illicit flows, with North Korean-linked addresses laundering an estimated $51 billion in 2024 [33]. For each gateway, it is important to integrate a real-time blockchain-analytics oracle that:
- Screens inbound/outbound addresses against Office of Foreign Assets Control (OFAC), EU, and UN lists
- Assigns a risk score derived from cluster analysis of darknet service exposure
- Auto-quarantines pending enhanced due diligence to prevent the risk of transactions exceeding a pre-set threshold.

### 7.6 Operational considerations

Running a single tokenized-dollar ledger across several legal domains is as much an engineering exercise as it is a regulatory one. Even when AML and data-privacy rules are satisfied, the day-to-day requirements of settlement, messaging, and governance still present a distinct layer of operational friction.

The first challenge is message-format fragmentation. Every token transfer is visible through an on-chain block explorer, giving auditors a single JSON record of hash, amount, and timestamp. That visibility, however, does not eliminate the need to move reserves through legacy real-time-gross-settlement (RTGS) systems that speak different dialects: SWIFT MT-103 in the United States, ISO 20022 pacs.009 in the euro area, and a patchwork of domestic formats elsewhere. Each format expects a field settlement date, purpose code, and fee base that are not natively stored in a token transfer. The network therefore introduces a bridge layer that translates redemption instructions before they hit the chain, mapping them to a canonical ISO 20022 business-purpose code. The result is a single audit trail for supervisors, while issuers avoid maintaining bespoke "explorer-scrape to payment-message" adapters.

A second challenge arises from data-localization mandates. Countries such as India and the United Arab Emirates require that any ledger holding resident data be hosted on servers physically located within their borders. Replicating the chain inside those jurisdictions undoubtedly improves fault tolerance since more copies mean a lower probability of data loss or censorship, but it can also lead to "regulatory forks" if a local supervisor insists on running modified software or censoring sanctioned addresses at the protocol layer.

To balance resilience with consensus integrity, the design deploys regulatory tools: read-only validator nodes that mirror the canonical chain in real time yet lack the keys to write blocks. Supervisors gain full visibility and satisfy sovereignty rules, while the authoritative ordering of transactions remains global.

Around-the-clock token liquidity poses a third operational problem, because central-bank RTGS rails still shut down overnight and on weekends. Without careful design, a token redeemed in one time zone could over-draw the issuer's reserve account in another. The hybrid model therefore maintains a prefunded liquidity grid: omnibus reserve accounts in New York (USD), Frankfurt (EUR), and Singapore (SGD) are topped up each business morning, and smart-contract logic caps redemptions to the value already on deposit until the next RTGS window opens. Holders enjoy 24×7 convertibility; issuers never incur daylight-overdraft risk.

Finally, the system must accommodate coordinated smart-contract upgrades, whether to patch a security vulnerability or to implement a new regulatory rule. Upgrades follow a two-tier governance process: a two-thirds super-majority of licensed issuers must vote in favor, after which a 24-hour time-lock exposes the finalized code to all home supervisors before activation. Execution occurs through a proxy pattern, so the token balances themselves remain in an immutable contract while the upgradeable logic resides behind a replaceable pointer. That structure ensures continuity of legal title, minimizes downtime, and gives every regulator a chance to veto non-compliant code before it goes live.

Taken together, these mechanisms convert what could be a patchwork of bilateral fixes into a cohesive operational framework. They preserve the security benefits of global replication and on-chain transparency yet keep the system compatible with the heterogeneous settlement rails, sovereignty rules, and supervisory calendars that define today's monetary reality.

## 8 Liquidity risk analysis: baseline vs. hybrid monetary model

To quantify how a hybrid monetary model helps in mitigating liquidity risk relative to a baseline design, we have calculated a number of metrics to measure the improvements.

## 8.1 Data

We have collected minute-by-minute USDC-USD price on Kraken and Circle's historical daily USDC redemptions starting from 2023-3-10 00:00 UTC to 2023-03-15 00:00 UTC (during the SVB episode) as the data for the following calculations.

For reserve parameters, we have derived the following numbers from the Circle's nearest disclosed auditing reports: Cash in reserve 12%, T-bills 45%, repos 43% of float; USDC float (scaling) $43 billion. Also, for T-bill haircut, we applied 2% for ≤1 hour and 0% for 24 hours.

## 8.2 Metrics

*Funding liquidity* [34]: Let $F$ be the float of the stablecoin; $C, B$ be the cash and T-bill shares in the stablecoin reserves; $\alpha_c$ be the ≤1 hour cash access factor; $h_B$ be the ≤1 hour T-bill haircut (pledged via a standing line). Using daily redemptions data, we estimate the p-quantile of 24-hour outflows, $Q_{24h}(p)$, and map it to 1-hour via a worst hour share $\phi: Q_{1h}(p) = \phi Q_{24h}(p)$. The instantly monetizable reserves (IMR) within horizon $\Delta t$ are

$$IMR_{\Delta t} = \alpha_c CF + \sum_i (1 - h_i) Conv_i(\Delta t)$$

with $Conv_i(\Delta t)$ be the amount of asset $i$ convertible with in $\Delta t$. T-bills contribute $(1 - h_B)\min\{BF, line\ cap\}$ at ≤1 and $BF$ at 24 hours.

Coverage and shortfall are calculated as:

$$ILCR_{\Delta t} = \frac{IMR_{\Delta t}}{Q_{\Delta t}(p)}, \quad MMG_{\Delta t} = \max\{0, Q_{\Delta t}(p) - IMR_{\Delta t}\}$$

$ILCR_{1h}$ tests if the stablecoin model, whether current, hybrid, or any future ones, is able to clear 1-hour redemptions at par under the most stressed time frame. $ILCR_{24h}$ reflects T+1 settlement. $MMG_{1h} = 0$ is the deign goal (no short fall).

For the parameters being used (SVB-calibrated, calculated based on the redemption flow of USDC provided by Tokenterminal.com): $F = \$43$ billion; $C = 0.12$; $B = 0.45$; $\alpha_c = 0.50$; $h_B = 0.02$; $p = 0.99$; $\phi = 0.75$.

*Market (peg) model and persistence metrics* [35]: With 1-minute mid-price $P_t$, we define the baseline deviation as $D_t = 10,000|P_t - 1|$ (bps), and summarize:

$$D_{max} = \max D_t, \quad M_\varepsilon = \sum_t \mathbb{1}\{D_t \geq \varepsilon\}, \quad L_\gamma = \max\{consecutive\ run\ length\ D_t \geq \gamma\}$$

$M_\varepsilon$ indicates the total minutes outside a $\pm\varepsilon$ band (we use $\varepsilon = 5bps$ in our result) and $L_\gamma$ refers to the longest run of stressed minutes (we use $\gamma = 10\ bps$ in our result).

To model the hybrid design, we convert a par-redemption rail of capacity R (USD/min)

into effective depth [36] using a conservative pass-through parameter $\alpha \in (0, 1]$:

$$D_t^{hyp} = D_t^{base} \cdot \max\{\underline{s}, \frac{V_t^{eff}}{V_t^{eff} + \alpha R}\}, \quad V_t^{eff} = \max\{V_t, Vol_{Floor}\}$$

We set $Vol_{Floor}$ = \$5 million/min, the minimum scaling factor $\underline{s} = 0.25$, and $\alpha = 0.5$ by default and report $\alpha$ sensitivity. The same persistence metrics $(D_{max}, M_\varepsilon, L_\gamma)$ are recomputed on $D_t^{hyp}$. In this assessment, $\alpha$ is the realized share of rail capacity that behaves like tradable depth at the minute frequency after routing and latency. While $\alpha = 1$ is the perfect pass-through, $\alpha < 1$ is conservative. We then calibrate $\alpha$ empirically when possible.

*Operations (wait time) metrics* [37]: Let $\lambda$ be the worst hour request arrival rate. Derived from the 1-hour rail tail and an average ticket size S, $\lambda = (\frac{Q_{1h}(p)}{S})/60$ (req/min). With per-server service rate $\mu$ and $c$ servers, utilization preserver is $\rho = \frac{\lambda}{c\mu} < 1$. The Erlang-C wait probability and expected queuing wait (minutes) are:

$$P(wait) = \frac{\frac{a^c}{c!(1-\rho)}}{\sum_{k=0}^{c-1}\frac{a^k}{k!} + \frac{a^c}{c!(1-\rho)}}, \quad a = \frac{\lambda}{\mu}, \quad W_q = \frac{P(wait)}{c\mu - \lambda}$$

reported in seconds. The minimum servers to meet an SLA $\tau$ seconds is

$$c^* = \min\{c \in \mathbb{N}: 60W_q(\lambda, \mu, c) \leq \tau\}$$

### 8.3 Result and Analysis

The time-series plot in Fig. 3 shows that the counterfactual hybrid design materially attenuates both the amplitude and the persistence of the SVB-week de-peg. At the onset of stress, the baseline series exhibits a sharp overshoot, with minute-level deviations briefly exceeding 12% before settling into a prolonged plateau of several hundred basis points.

In contrast, the hybrid series caps the peak at roughly 3% and returns toward parity much faster, with visibly shorter excursions above the zero-deviation line. In other words, the hybrid compresses the height of the shock (≈ 75% reduction in the observed peak) and the time under stress (fewer and shorter runs outside a tight band), implying stronger mean-reversion of the peg. This pattern is consistent with added, immediately available depth at par: when order-book liquidity thins, the rail converts redemption demand that would otherwise push price away from \$1 into off-book settlement at par. Note that the standing line against T-bills underwrites that flow, so the market does not need to clear via price.

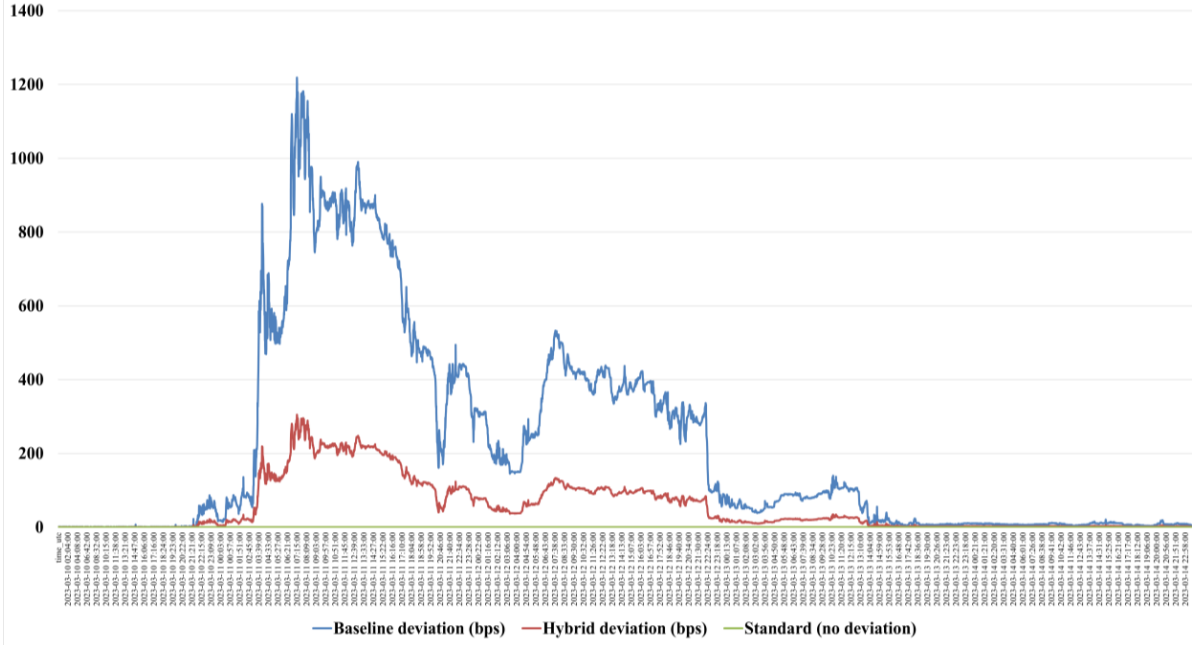

**Fig. 3** Peg deviation during the SVB crisis.

Table 1 reports the outflow tails used to size stress. Over the SVB-calibrated window, the p99 daily redemption is $1.849 billion, and we map day-to-hour using a worst-hour share of 75%, giving a p99 1-hour proxy of $1.387 billion. Relative to a $43 billion float, these correspond to about 4.3% of float in one day and 3.2% in the peak hour, which is a severe but historically grounded shock. The SVB-calibrated tails are larger than those from the full sample, so all comparisons are conservative.

**Table 1** Outflows

| Window | P95 24h (USD) | P99 24h (USD) | 1h proxy at 75% (USD) |
| --- | --- | --- | --- |
| Full sample | 906,858,640 | 1,525,993,255 | 1,144,494,941 |
| SVB-calibrated | 1,276,681,615 | 1,848,824,810 | 1,386,618,608 |

Table 2 shows the result of the metrics for both models. In the table, both scenarios report $ILCR_{1h} = 1.861$ and $ILCR_{24h} = 13.257$ with $MMG_{1h} = 0$. The equality across scenarios is by design: the reserve composition is held constant, so the hybrid does not change how much liquidity exists, but it changes how it is accessed and routed (market/operation channels).

**Table 2** Baseline model vs. hybrid model metrics outcome.

| Metric | Baseline | Hybrid | ΔHybrid | Δ% |
|---|---|---|---|---|
| $ILCR_{1h}$ | 1.861 | 1.861 | 0 | 0 |
| $ILCR_{24h}$ | 13.257 | 13.257 | 0 | 0 |
| $MMG_{1h}$ | 0 | 0 | - | - |
| Max peg deviation | 1219 | 304.8 | -914.2 | -75% |
| Peak wait time | ∞ | 57.7 | Stabilized | - |
| Minutes ≥ 5 bps (min) | 5442 | 3803 | -1639 | -30% |
| Longest run ≥ 10 bps (min) | 3799 | 2931 | -868 | -22.9% |

These ILCR numbers means that at the 1-hour horizon, the observed stress draw (≈$1.39 billion) is comfortably below the short-horizon liquidity that is operationally accessible under our assumptions (hence $ILCR_{1h} > 1$ and $MMG_{1h}= 0$ in both cases). Put differently, the data-implied "bad hour" never exhausts the first-hour liquidity bucket in this calibration. At the 24-hour horizon, the p99 day (≈$1.85 billion) is tiny relative to day-settling liquid assets (cash + T-bills), which is why the reported $ILCR_{24h}$ is an order of magnitude greater than one (13.257) for both scenarios. This reflects a straightforward empirical comparison, which the size of the realized worst-day tail is far below the size of assets that settle by T+1.

The hybrid materially improved the condition of the de-peg under simulation. According to the table, the hybrid system shows a reduction in maximum peg deviation from 1219 bps to 304.8 bps (Δ = −914.2 bps; −75%). Persistence also improves: de-peg time under stress (minutes with deviation ≥ 5 bps) falls from 5,442 to 3,803 (Δ = −1,639; −30%), and the longest run time (longest consecutive stretch ≥ 10 bps) declines from 3,799 to 2,931 minutes (Δ = −868; −22.9%). These changes are economically meaningful because they indicate both a lower crest and faster mean reversion of price back to $1, consistent with the hybrid system's par-redemption rail and collateralized access to T-bill liquidity providing immediate depth at par precisely when order-book liquidity thins.

On the operational side, the baseline queue is unstable at the worst hour (reported as ∞ in Table 3), while the hybrid system stabilizes the situation with a bounded peak expected wait of 57.7 seconds. This result is consistent with the implied arrival intensity: the p99 1-hour flow of $1.387 billion with a $1 million average ticket corresponds to roughly 1,387 requests in an hour, i.e., ~23 requests/minute. With a service rate of 2 req/min/server, baseline capacity (5 servers) is only 10 req/min (utilization ≥ 1 means unstable), whereas hybrid capacity (12 servers) is 24 requests/minute (utilization just below 1), yielding the reported ~58 seconds expected wait via Erlang-C.

Taken together, the results show that under an SVB-calibrated shock the hybrid monetary design improves exactly where run dynamics propagate. It cuts peak price dislocations by three-quarters, reduces the duration and streakiness of stressed pricing, and keeps redemption lines moving, which all without changing the reserve mix.

## 9 Challenges and considerations

The implementation of a hybrid monetary ecosystem with stablecoins faces several regulatory and legal challenges, most of which are ultimately tied to liquidity management under stress. Current regulatory divergences due to different classifications of stablecoins as e-money, securities, or commodities across jurisdictions, create uncertainty regarding which entities can act as liquidity backstops and how redemption flows are prioritized in a stress event.

In the United States, for example, the SEC has indicated that some stablecoins may constitute securities [38], while the CFTC and the Treasury have taken different positions [39]. This lack of definitional consensus complicates efforts to integrate stablecoins into an applicable national structure and slow down innovation due to uncertainty over which agencies have oversight authority. In addition, the integration of private stablecoins with public infrastructure raises questions around consumer protection and systemic risk. A key concern is ensuring that issuers maintain full reserve backing and do not engage in risky lending, as occurred during the 2022 Terra/LUNA collapse [40]. Even fiat-backed stablecoins like USDT have faced criticism over its reserve disclosures.

A second challenge arises from ensuring that liquidity buffers are both sufficient and accessible. Newly adopted U.S. legislations, such as the Stablecoin TRUST Act 2022 and the GENIUS Act 2025, seek to establish regulatory clarity by requiring stablecoin issuers to hold reserves in cash or government securities and submit to periodic audits, but these requirements alone do not guarantee immediate redemption capacity. Liquidity is not only a matter of collateral quality but also of collateral convertibility. That is, whether reserves can be mobilized intraday without haircut or delay. Regulatory frameworks must therefore explicitly define permissible reserve assets, redemption rail obligations, and access to central bank liquidity facilities, particularly under stressed market conditions.

Cross-border regulatory inconsistencies compound these risks. Stablecoins are inherently borderless, and liquidity stress can transmit across jurisdictions through arbitrage flows and redemption cascades. Existing AML/CFT and Travel Rule requirements provide important guardrails but can also introduce frictions that slow liquidity mobilization [41]. A coordinated regulatory framework must therefore ensure that compliance protocols do not inadvertently obstruct rapid redemption capacity, which is critical during a run.

Finally, supervisory regimes must adapt to the unique nature of liquidity dynamics in tokenized environments. Stablecoins can experience redemption spikes that are orders of magnitude faster than traditional banking withdrawals. Consequently, oversight should focus not only on solvency or collateral sufficiency but also on real-time liquidity monitoring, interoperability of backstops, and pre-committed standing facilities. This liquidity-centric regulatory architecture would align operational capabilities with systemic stability goals.

The idea of creating a hybrid monetary ecosystem with a public-private partnership to balance oversight and innovation has been supported by many industry practitioners and academic researchers. In particular, many studies have explored how stablecoins and traditional financial institutions can coexist and evolve together using a two-tier architecture in which the retail layer is for user-level transaction while the wholesale layer is for interbank settlements using CBDCs for liquidity and stability [42]. With such a hybrid system, banks are not displaced but repurposed as on/off ramps between fiat and stablecoins. They could also issue tokenized deposits and offer interest-bearing digital assets while performing their important role in regulatory compliance of AML/KYC.

## 10 Conclusion

Stablecoins are transitioning from niche financial instruments to foundational elements of global monetary infrastructure. As their scale expands, liquidity risk emerges as the single most consequential driver of systemic vulnerability. This paper develops a hybrid monetary architecture designed to neutralize such risk through structural features.

By embedding fiat-backed stablecoins within a central bank–anchored framework with 100% reserve backing, interoperable redemption rails, and standing liquidity lines, the model directly targets the mechanisms through which liquidity crises propagate, namely, delayed redemption, maturity-transformation fragility, and fragmented settlement infrastructures. Our SVB–USDC stress test demonstrates how this approach can materially reduce peak peg deviations and stress persistence, offering a quantifiable liquidity stabilization effect without altering reserve composition.

More broadly, this liquidity-focused architecture provides a credible path toward integrating stablecoins into regulated monetary systems while preserving programmability, financial innovation, and monetary integrity. By treating liquidity as a design principle rather than an operational afterthought, regulators and market participants can build a system in which stablecoin stability is assured by structure, not just policy response.